\documentclass{ics}
\usepackage{color}
\usepackage{fullpage}
\usepackage{mytex}
\usepackage{newtype}

\newtheorem{theorem}{Theorem}

\newtheorem{lemma}[theorem]{Lemma}

\begin{document}

\title{Bounding Rationality by Discounting Time}

\author{
  Lance Fortnow$^{1}$
  \and
  Rahul Santhanam$^{2}$}

\address{
  $^{1}$Northwestern University, Evanston, USA
  \and
  $^{2}$University of Edinburgh, Edinburgh, UK}

\email{
  fortnow@eecs.northwestern.edu
  \and
  rsanthan@inf.ed.ac.uk}

%\author{Lance Fortnow\thanks{Supported in part by NSF grants CCF-0829754 and DMS-0652521.}\\Northwestern University\\{\normalsize \tt fortnow@eecs.northwestern.edu}
%\and Rahul Santhanam\\University of Edinburgh \\{\normalsize \tt
%rsanthan@inf.ed.ac.uk}} \maketitle

\begin{abstract}
Consider a game where Alice generates an integer and Bob wins
if he can factor that integer. Traditional game theory tells us
that Bob will always win this game even though in practice
Alice will win given our usual assumptions about the hardness
of factoring.

We define a new notion of bounded rationality, where the
payoffs of players are discounted by the computation time they
take to produce their actions. We use this notion to give a
direct correspondence between the existence of equilibria where
Alice has a winning strategy and the hardness of factoring.
Namely, under a natural assumption on the discount rates, there
is an equilibrium where Alice has a winning strategy iff there
is a linear-time samplable distribution with respect to which
Factoring is hard on average.

We also give general results for discounted games over
countable action spaces, including showing that any game with
bounded and computable payoffs has an equilibrium in our model,
even if each player is allowed a {\it countable} number of
actions. It follows, for example, that the Largest Integer game
has an equilibrium in our model though it has no Nash
equilibria or $\epsilon$-Nash equilibria.

\end{abstract}


\maketitle

\section{Introduction}

Game theory studies the strategic behavior of self-interested
rational agents when they interact. In the traditional setting
of game theory, agents are supposed to be perfectly rational,
in terms of knowing what their strategic options and the
consequences of choosing these options are, as well as being
able to model perfectly the rationality of other agents with
whom they interact. However, often in practice, when human
beings are involved in a strategic game-playing situation, they
fail to make perfectly rational decisions. Herbert Simon first
developed this ``bounded rationality'' perspective.

In the past couple of decades various models of bounded
rationality~\cite{Neyman85, Abreu-Rubinstein88, Binmore88,
Megiddo-Wigderson86, Papadimitriou-Yannakakis94,
Fortnow-Whang94} have been defined and studied by game
theorists and computer scientists. In this paper, we introduce
a new notion of bounded rationality based on the perspective of
computational complexity. We argue that it is natural, and
prove that it has some nice properties and can be used to
obtain new connections between game theory and computational
complexity.

The main idea is to discount the payoffs of players in a game
based on how much time they take to play their actions, with
different players possibly discounted at different rates. Of
course, we need to define what it means for a player to take
time to play its action. This naturally pre-supposes that each
player has some computational mechanism for playing its
strategy - in this paper, as in the recent work by Halpern and
Pass\cite{Halpern-Pass08}, we adopt the probabilistic Turing
machine as our computational model. This is a computational
model which is universal, and is also generally considered to
be realizable in Nature. Furthermore, it capture complexity via
running time and can be used to realize games with countable
action spaces, unlike say if we were to use finite automata:
the model typically considered by game theorists when studying
bounded rationality.

In this paper, we use {\it exponential} discounting, meaning
that the payoff goes down by a factor $(1 - \delta)^t$ after
time $t$, where $\delta$ is a constant. Our main results also
hold for other notions of discounting, as we discuss in Section
\ref{discussionsec}.

The notion of discounting is far from new \cite{Samuelson37,
Koopmans60, Frederick-Loewenstein-O'Donoghue04} - indeed much
of economic theory depends on it. It is a basic economic
assumption that people value a dollar a year from now less than
a dollar today. The discount $1-\delta$ for a specific period
is chosen so that an agent is indifferent between receiving
$1-\delta$ dollars now and 1 dollar at the end of the period.

Discounting is commonly used for computing cumulative payoffs
in repeated games. We emphasize that we discount based on {\it
computation} time, which means that the notion of discounting
can now even be used for one-shot games even when there is no
natural notion of input size. The idea of discounting based on
computation time was developed by Fortnow~\cite{Fortnow08},
where he used it for a variaton on the ``program equilibria''
framework devloped by Tennenholtz~\cite{Tennenholtz04};
moreover, a single discount rate is used for all players.

Our notion of discounted time has several benefits. First, it
bounds rationality {\it endogenously} rather than exogenously.
By this, we mean that the bound on an agent's rationality is
not imposed from outside, but rather arises from the agent's
own need to maximize its utility.

Second, discounting has some nice mathematical properties. It's
time-independent - discounting for $r$ steps starting at a time
$t_0$ yields the same relative decrease in payoff as
discounting for $r$ steps starting at an earlier or later time.
Given a discount factor $1 - \delta$, the discounted payoff
behaves like a linear function for small $t$ and like an
exponential function for large $t$, which accords well with our
intuitions for how we value computational resources in the real
world. We might only be marginally more gratified by a
computational task finishing in 1 second than one finishing in
2 seconds, but we would certainly be far more annoyed if a task
finished in 20 minutes than in 10 minutes.

Also, the discounting model is philosophically elegant in that
it {\it unifies} time as viewed by economists and time as
viewed by computer scientists. Time is an important concept
both in economics and in computational complexity, and we model
it in a way that is consistent with the perspectives of both
fields.

We use asymmetric discounting in our model - different players
may have different discount factors. There are a couple of
reasons for this. First, players might have asymmetric roles in
a game, and in this case it is natural to give them discount
factors. For example, a cryptographic protocol can be
interpreted as a game where players are either honest or
adversarial. In this setting, it makes sense to model the
adversary as more patient and therefore having discount rate
$\delta$ closer to 1.

However, even if all the players are equally patient with
respect to real time, it still makes sense to give them
different discount factors. This is because discounting is done
as a function of {\it computational} time rather than real
time, and the relationship between computational time and real
time depends on the power of technology. If one player has a
much faster computer than the others, then it is effectively
more patient, in that it has a smaller discount factor. For
example, consider a two-player game where the players are
equally patient in that the payoff for each player halves after
1 second of real time. Suppose, however, that Player 1 has a
computer with a clock rate of $10^{6}$ operations per second,
and Player 2 has a computer with a clock rate of $10^{12}$
operations per second. Then the discount rate $\delta_1$ for
Player 1 is approximately $10^{-6}$ and the discount rate
$\delta_2$ for Player 2 is approximately $10^{-12}$.

This is a further advantage of our model, in that it factors in
the power of technology. Many games today play out in a virtual
setting, eg. the game between someone sending their credit card
information and a malicious adversary seeking to steal their
identity, or an electronic auction, or even computer chess. In
all these cases, the power of technology has a critical impact
on strategy and success in the game, which is not modeled
adequately by traditional game theory. Not only do we model
this via the discount rates, but our notion of uniform
equilibrium also implicitly models how technology {\it evolves}
with time.

Our model exhibits some nice phenomena for general classes of
games. We define a new notion of equilibrium for our model,
which we call ``uniform equilibrium''. We show that for finite
games, there's a uniform equilibrium corresponding to every
Nash equilibrium. For games where each player has a countable
action space, the situation is even more interesting. It's
known that Nash equilibria do not exist in general in this
case. However, under mild assumptions, namely that the payoffs
are bounded and computable, we show that uniform equilibria
{\it always} exist even in this case.

As an example, consider the Largest Integer game, where each
player outputs a number and the player outputting the largest
number wins the entire pot of money at stake (with the players
sharing the pot equally if they output the same number). This
is an archetypal example of a game which has no Nash equilibria
or even approximate Nash equilibria. The absence of Nash
equilibria means that traditional game theory provides no
predictive or explanatory framework for how the game will
actually play out.

The Largest Integer game does have a uniform equilibrium in our
framework, and there is an intuitive explanation of this.
Essentially, the Largest Integer game models oneupmanship,
where each player is trying to outdo the other. What is not
modeled by traditional game theory is that the players expend
considerable resources in this process, which affects their
``effective payoff''. Indeed, as more and more resources are
required, at some point the players become essentially
indifferent between winning and losing. In our case, the
resource is time; the equilibrium situation corresponds to both
players spending so much time coming up with and writing down a
large number that their payoffs are driven to zero by their
discount factors.

\subsection{The Factoring Game}

Perhaps the most interesting results in this paper concern a
close relationship between equilibria in discounted games and
the computational complexity of problems. We illustrate this
using the Factoring game.

The Factoring game is a puzzle in the theory of bounded
rationality. Consider the following game between two players
Alice and Bob. Alice sends an integer $n \geq 2$ to Bob, who
attempts to find its prime factorization. If Bob succeeds, he
``wins'' - he gets a large payoff and Alice gets a small
payoff; if he fails, the opposite happens.

If formulated as a game in the conventional way, Bob always has
a winning strategy. However, in practice, one would expect
Alice to win, since factoring is believed to be computationally
hard. This is the puzzle: to find a natural formulation of the
game that captures the intuition that Alice should win if
factoring is indeed computationally hard.

The Factoring game was first introduced by Ben-Sasson, Kalai
and Kalai ~\cite{BenSasson-Kalai-Kalai06} and also considered
by Halpern and Pass~\cite{Halpern-Pass08}. Neither gives an
explicit solution to the puzzle, instead they give general
frameworks in which to study games with computational costs.
Indeed, Ben-Sasson, Kalai and Kalai say in the Future Work
section of their paper that ``it would be interesting to make
connections between asymptotic algorithmic complexity and
games''.

We show that the structure of equilibrium payoffs in the
discounted time version of the game corresponds closely to the
computational complexity of factoring. Specifically, if
Factoring is in probabilistic polynomial time on average, Bob
always wins; if not, there are equilibria in which Alice gets a
large payoff. This result assumes that the discount rates of
the two players are polynomially related - we motivate this
assumption in Section \ref{factoringsec}. If there's a
different relationship between the discount rates, then there's
a corresponding different complexity assumption which
characterizes when Alice has a winning strategy. In the
simplest interpretation of our model, where discount rates are
determined by the power of technology, it can be empirically
tested how discount rates vary with each other.

What makes this connection with asymptotic complexity somewhat
surprising is that the notion of input length is not explicitly
present in our model. Instead, it arises naturally from the
discounting criterion and our notion of uniform equilibrium.

The Factoring game is relevant not only to game theory, but
also to the foundations of cryptography. There has been a lot
of research into the connections between game theory and
cryptography \cite{Dodis-Rabin07,Katz08}, but much of this has
focused on multi-party computation.
 One can define an analogue of the Factoring game for any
one-way function and obtain similar results; there's nothing
special about Factoring being used in the proofs. This
game-theoretic perspective might be useful in studying the
tradeoff between efficiency of encryption and security in
cryptosystems. In general, it would be interesting to
investigate a perspective where the success of a cryptosystem
depends on the adversary being ``bounded rational'' rather than
computationally bounded in some specific sense.

\subsection{Further Discussion of the Model}
\label{discussionsec}

Here, we further discuss various features of our model and
compare it to alternative ones.

Our criteria for a reasonable model is that it should be {\it
general}, i.e., be relevant to a class of situations rather
than a single specific situation, and that it should have {\it
explanatory power}, i.e., not only should it simply correspond
to an observed phenomenon but provide some further insight. For
comparative purposes, in the context of the Factoring game, one
can think of some alternative models that predict a win for
Alice. For example, one could imagine that the players have a
fixed finite amount of time to make a decision, with Alice
given say 10 seconds to choose her number, and Bob 100 seconds
to respond with the prime factors. It's clear that if Bob can't
factor a random large number (which could be generated quickly
by Alice), he loses, however this is an unsatisfactory model in
many respects. First, it deals with a very specific situation,
so it cannot say anything about computational complexity or how
equilibria depend on the power of technology. Second, the model
is inherently non-robust. Bob might be able to factor Alice's
number in 101 seconds - in a real-life situation, this
difference shouldn't affect his payoff too much, but in this
model, it does. By adopting a {\it flexible} model of bounded
rationality, where payoffs degrade continuously with time, we
avoid such pathological effects.

One way to make the fixed-time model more general is  quantify
over the time limit: to say, for example, that if Alice is
allowed $t$ units of time, then Bob is allowed $t^2$ units of
time. This kind of approach is taken when formulating the
notion of ``computational equilibrium'' \cite{Urbano-Vila02,
Dodis-Rabin07} where they limit the set of machines being used
to those that run in some security parameter where our model
makes no such restriction on machines but control time with
utility. Another problem with the computational equilibrium
model is that though it might be consistent with the observed
phenomenon, it's unclear {\it why} the assumptions the model
makes should hold. In such a case, the model is simply a way to
re-formulate a phenomenon, rather than an explanation for it.
In contrast, in our model, there are clear motivations for the
choices made. Discounting is based on time preference of
utility, which is well established and extensively used in
economics \cite{Frederick-Loewenstein-O'Donoghue04}. Also our
interpretation of discount rates in terms of the power of
current technology matches the intuition that a player armed
with a more powerful computer should be able to make a more
rational decision, i.e., more in its self-interest. Finally,
our use of asymmetric discount rates models asymmetries in the
roles of players and in the power of technology available to
them.

Regarding some of the more specific choices made, one could
question why we use exponential discounting rather than some
other form of discounting. Exponential discounting is still the
discounting model of choice in economics and game theory, but
there have been arguments made that other models such as
``hyperbolic discounting'' more accurately represent human time
preference of
utility~\cite{Frederick-Loewenstein-O'Donoghue04}. As it turns
out, the exact choice of discounting model does not matter very
much to us - our main results on the Factoring game and the
general result on bounded-payoff games (Theorems
\ref{Player1Win}, \ref{Player2Win} and \ref{EqExist}) go
through even in the hyperbolic discounting model and, we
suspect, in any reasonable model of discounting.

Another issue which can be debated is whether each player's
utility is discounted only by its own computational time or by
some function of its computational time and the computational
time of the other players. In a strategic situation, it seems
natural to penalize a player only for its own computation.
Consider a two-player simultaneous-move game, where each player
plays without knowledge of the other player's action. Suppose
Player 1 plays first in this game. Should Player 1 be charged
for Player 2's time as well, since the outcome is determined
only after Player 2 has played? We think not, since Player 1
can use its extra time doing other things, garnering utility in
other ways. Of course, if one player plays first, that might
seem to ``sequentialize'' the game. For our model to apply,
there has to be a mechanism in place to ensure that the players
do in fact act independently.

Our model can, in principle, deal with both positive and
negative payoffs - discounting predicts, as seems intuitive,
that positive payoffs should motivate agents to play quickly,
while negative payoffs should cause agents to procrastinate.
However, in this paper, we deal only with games with positive
payoffs. This is because it's tricky to define what happens if
the first player's computation finishes within a finite time
but the second player's strategy computation never halts. In
some sense, this corresponds to the second player not playing
the game at all. With strictly positive payoffs, we can be
guaranteed that in an equilibrium situation, all players will
play within a finite time - it is in the interest of all
players to play as quickly as possible. A way to avoid the
issue with positivity of payoffs would be to give players
preference orderings on outcomes rather than ascribing real
payoffs, as is often done in game theory
\cite{Osborne-Rubinstein94}, and have the preference orderings
vary with computational time. Though perhaps a more accurate
model, this has the disadvantage of being very cumbersome
mathematically.

\subsection{Related Work}

Bounded rationality is a rich area, with lots of work in the
past couple of decades. We survey some of that work and clarify
the relationship to our ideas, with an emphasis on more recent
work. There are several excellent surveys and references on
bounded rationality \cite{Aumann92, Binmore88, Kalai90,
Rubinstein98}.

Early work focused mainly on bounded rationality in the context
of the repeated Prisoner's Dilemma game, where strategies are
modeled as finite automata \cite{Abreu-Rubinstein88,
Gilboa-Samet89, Neyman85, Papadimitriou-Yannakakis94}. There
were some works during this period which modeled strategies by
Turing machines \cite{Megiddo86, Megiddo-Wigderson86}, but
these works were concerned with Turing machine size as a
complexity measure rather than time. There has also been a good
deal of work in the economics literature studying the
consequences for economics of the constraint that agents act in
{\it computable} ways \cite{Canning88, Binmore88, Norman94},
but these works do not deal with computational complexity.

Recently there has been a resurgence of interest in modeling
strategies as general Turing machines. We note especially the
two papers \cite{BenSasson-Kalai-Kalai06, Halpern-Pass08} which
discuss the Factoring game. Rather than specifying an explicit
solution to the puzzle of the Factoring game, these works
provide general frameworks and results for taking computational
costs into account when playing games. Our contribution in this
paper is in providing a concrete and natural model which
captures the cost of computational time, and using it to solve
the Factoring puzzle.

Other recent works \cite{Tennenholtz04, Fortnow08} consider
computer programs as strategies, but in the context of a
different kind of equilibrium known as the program equilibrium,
where rationality is modeled by letting each player's program
have as input the code of the other player's program. As
mentioned earlier, Fortnow \cite{Fortnow08} considers
discounted computation time in this context to obtain a broader
range of program equilibria rather than to model bounded
rationality, and he allows only for a single discount factor.

The idea of discounting time has also been proposed in the
completely different context of verification
\cite{AlfaroHenzingerMajumdar03}.

\section{Preliminaries}
\label{prelimsec}

%We assume knowledge of standard concepts in complexity theory - the
%textbook by Arora and Barak \cite{Arora-Barak09} is a good
%reference.

We review standard concepts for two-player games. For a more
detailed treatment, refer to the books by Osborne and
Rubinstein \cite{Osborne-Rubinstein94} and Leyton-Brown and
Shoham \cite{LeytonBrown-Shoham08}.

%
%We consider both strategic games and extensive games with countable
%strategy spaces. In a strategic game, neither player has knowledge
%of the other player's moves when making its move. In an extensive
%game, one of the players moves first, and the other player makes its
%move with knowledge of the first player's move. We may assume
%without loss of generality that Player 1 moves first.

In this paper, we only consider {\it one-shot} games of {\it
perfect information}, where each player makes a single move. We
represent these games in {\it normal form} as a four tuple $G =
(A_1, A_2, u_1, u_2)$, where $A_i$ is the action space for
player $i$. The utility function $u_i: A_1 \times A_2
\rightarrow \Re^{\geq 0}$ is a payoff function specifying the
payoff that accrues to player $i$ depending on the actions
played by the two players. We consider both the simultaneous
version where both players play their actions simultaneously
and the sequential version where player 2 can base his action
on the action taken by player 1.

As mentioned before, we assume in this paper that payoff
functions are always non-negative.

Strategies describe how the player's choose their actions. A
pure strategy for Player 1 is simply an element of $A_1$. For
simultaneous-move games, a pure strategy for player 2 is just
an element of $A_2$. For sequential games, a pure strategy for
player 2 is a function from $A_1$ into $A_2$. We use $S_i$ to
represent the pure strategy space for player $i$ and we extend
the utility functions $u_i$ to strategies in the natural way.

A mixed strategy for a player is a probability distribution
over its pure strategies. The payoff for a game using the mixed
strategies is just the expected payoff when each player chooses
their strategies independently from their chosen distributions.

%When we say ``strategy'' in this paper, we mean a pure strategy by
%default; we try to be more explicit when there is a probable
%ambiguity.

%We denote the class of probability distributions over strategies for Player
%$i$ by $\Delta(A_i)$.
%The payoff function $u_i$ can be extended to a partial function $U_i: \Delta(A_1) \times \Delta(A_2) \rightarrow \Re^{\geq 0}$, where for each mixed
%strategy pair $(S,T)$, $U_i(S,T)$ is the expectation over the probability
%distributions $S$ and $T$ of $u_i(s,t)$, where $s \in A_1$ and $t \in A_2$.
%Note that this expectation need not always be finite, even if each individual
%payoff is finite.
%
%We often abuse notation and use $u_i$ to refer to payoffs even for mixed
%strategy pairs - it will be clear from the context what we mean.

A pure-strategy Nash equilibrium (NE) is a pair of strategies
$(s_1, s_2) \in S_1
 \times S_2$ such
that for any $s^*_1 \in S_1$ and $s^*_2 \in S_2$, $u_1(s_1,
s_2) \geq u_1(s^*_1,s_2)$ and $u_2(s_1, s_2) \geq u_2(s_1,
s^*_2)$. A pair of strategies is an $\eta$-NE if neither player
can increase its payoff by more than $\eta$ by playing a
different strategy, given that their opponent plays the same
strategy as before. For small $\eta$, the players might be
satisfied with an $\eta$-NE rather than a pure NE, since they
might be indifferent to small changes in their payoff function.

A mixed-strategy Nash equilibrium is a pair of mixed strategies
for which neither player can increase their expected payoff by
playing a different mixed strategy, assuming that their
opponent plays the same mixed strategy as before. The notion of
an $\eta$-NE for mixed strategies is defined in an analogous
way to the definition for pure strategies.

The famous theorem of Nash \cite{Nash51} states that every game
over compact action spaces has a mixed-strategy Nash
equilibrium. When we say ``Nash equilibrium'' in this paper, we
mean a mixed-strategy Nash equilibrium unless otherwise stated.

\section{Our Model}
\label{modelsec}

The normal-form representation of a game does not say anything
about how a strategy is actually implemented by a player.
Depending on the method of implementation used, there might be
further costs incurred - the analysis of these costs may itself
be game-theoretic. This insight is formalized by the notion of
a {\it metagame}. Given a game $G$, the metagame is a new game
which augments $G$ by modeling outside factors which are
relevant to playing $G$. Thus a metagame aims to be a more
accurate model of how $G$ might play out in the real world.

We consider the {\it machine metagame}, which presumes that a
strategy is implemented by some computational process. We model
the computational process as a probabilistic Turing machine,
which is a very general model of computation. By the
Church-Turing thesis, probabilistic Turing machines can compute
any function that is effectively computable. The motivation for
considering probabilistic machines is the idea that randomness
is also a resource available in the real world.

In the machine metagame corresponding to a game $G = (A_1, A_2,
u_1, u_2)$, actions for Player $i$ are probabilistic Turing
machines rather than elements of $A_i$. Since we only consider
countable strategy sets, for each $i$ the elements of $A_i$ may
be represented by binary strings in some canonical way, with
each string representing a strategy and each strategy
represented by a string. If a probabilistic TM played by Player
1 outputs a string $x$ with probability $p(x)$, this is
interpreted as Player 1 playing a strategy $x$ with probability
$p(x)$ in the game $G$.

Now that strategies are Turing machines, computational issues
can be factored into the game, even though for a fixed game,
there is no natural notion of an ``input size.'' We address
this issue by discounting each player's payoff by the time
taken to produce a (representation of a) strategy. The discount
factors for the two players might be different, reflecting the
possibilities that the game is asymmetric between the two
players, and that the two players have differing amounts of
computational resources.

Given a game $G = (A_1, A_2, u_1, u_2)$, we formally define the
$(\epsilon, \delta)$-discounted version of $G$. This is the
discounted time machine metagame corresponding to $G$, where
the player's computation times are discounted by $1 - \epsilon$
and $1 - \delta$ respectively. In this game, each player's
action space is the class of all probabilistic Turing machines.
Each player's Turing machine gets as input $\lceil 1/\epsilon
\rceil$ and $\lceil 1/\delta \rceil$ in binary - this
corresponds to the players having full information about the
game. If the game is extensive, Player 2's Turing machine gets
as additional input the output of Player 1's Turing machine.

We formally specify how payoffs are determined. We first
consider the case where both player's Turing machines halt on
all computation paths. Given a computation path $z$ of a
probabilistic TM, let $t(z)$ denote the length of the
computation path (i.e., the time taken by the computation),
$f_1(z) \in A_1$ the action in $A_1$ corresponding to the
output of the path $z$, and $f_2(z) \in A_2$ the action in
$A_2$ corresponding to the output of the path $z$. Then the
payoff $u_1(M,N)$ of Player 1 corresponding to Player 1 playing
a probabilistic Turing machine $M$ and Player 2 playing $N$ is
the expectation over computation paths $z$ and $w$ of $M$ and
$N$ respectively of $(1-\epsilon)^{t_z}u_1(f_1(z), f_2(w))$.
Similarly, the payoff $u_2(M,N)$ of Player 2 is the expectation
of $(1-\delta)^{t_w}u_2(f_1(z),f_2(w))$.

In addition, we require a convention for payoffs on non-halting
paths. In this case, a player whose machine does not halt gets
payoff 0 (corresponding to discounting for infinite time), and
if the other player's machine does halt, the player gets the
maximum possible payoff over all actions in $A_1$ of playing
its action, discounted by the computation time of playing its
action.

%We have specified payoffs for pure stategy pairs in the discounted
%game above. These payoff functions are extended to payoff functions
%for mixed strategy pairs in the usual way. We again abuse notation
%and use $u_1$ and $u_2$ for the payoff functions of the two players
%in the discounted game - it will be clear from the context what is
%meant.

We define two new equilibrium concepts, which correspond to
equilibria that are robust when the discount rates $\epsilon$
and $\delta$ tend to zero. Our motivation for being interested
in this limiting case is that computational costs grow smaller
and smaller with time (or equivalently, computational power
increases with time) - this corresponds to $\epsilon$ and
$\delta$ approaching 0.

We say that a pair of probabilistic machines $(M,N)$ is a
uniform Nash equilibrium (NE) if for every pair of machines
$(M',N')$,
\[\liminf_{\epsilon,\delta\rightarrow 0} u_1(M,N)-u_1(M',N) \geq 0\]
and
\[\liminf_{\epsilon,\delta\rightarrow 0} u_2(M,N)-u_2(M,N') \geq 0.\]

%$max(0,) \rightarrow 0$ when $\epsilon, \delta \rightarrow 0$ and
%$max(0,u_2(M,N') - u_2(M,N)) \rightarrow 0$ when $\epsilon, \delta
%\rightarrow 0$.

We say that $(M,N)$ is a strong uniform NE of the discounted
game if there is a function $f$ such that $(M,N)$ is an
$f(\epsilon, \delta)$-NE for the $(\epsilon,
\delta)$-discounted game, for some function $f$ where
$f(\epsilon, \delta)$ tends to 0 when both $\epsilon$ and
$\delta$ tend to 0.

As the name indicates, the notion of a strong uniform NE is a
stronger concept since it requires a fixed equilibrium strategy
pair to be resilient in the limit against deviating strategies
which might depend on $\epsilon$ and $\delta$. In contrast, a
uniform NE is only required to be resilient in the limit
against other fixed strategies.

The definition of uniform equilibrium above assumes that
$\epsilon$ and $\delta$ are independent - i.e., the equilibrium
condition holds irrespective of how $\delta$ varies with
$\epsilon$, as long as they both tend to 0. In some of our
results, we will be concerned with the situation where $\delta$
is a function of $\epsilon$ such that $\delta \rightarrow 0$
when $\epsilon \rightarrow 0$. We will abuse notation by
referring to the corresponding notion of equilibrium, where the
limit is now taken only as $\epsilon \rightarrow 0$, also as a
uniform equilibrium.

We say that a payoff pair $(u,v)$ is a uniform equilibrium
payoff if there is a uniform equilibrium $(M,N)$ such that
$u_1(M,N) \rightarrow u$ and $u_2(M,N) \rightarrow v$ in the
discounted game when $\epsilon, \delta \rightarrow 0$

The above equilibrium concepts are defined for pure strategy
NEs, but the definitions extend easily to mixed strategy NEs.

%Occasionally, when we consider both a game $G$ and its discounted
%time version, we use the term ``action'' to refer to a pure strategy
%of the game $G$, to avoid confusion about what we mean by a strategy
%in this case.

All the definitions above can be generalized easily to
$N$-player games for $N > 2$ and indeed the results of the
Section~\ref{properties} all hold for $N$-player games as well.

\section{The Factoring Game}
\label{factoringsec}

In our formulation of the Factoring game, the winning player
receives a payoff of 2 (before discounting) and the losing
player receives a payoff of 1. The precise values of these
payoff are not important for our main results.

The $(\epsilon, \delta)$-discounted time version of the
Factoring game is defined in the usual way. In our presentation
here, we choose $\delta = \epsilon^c$, for some constant $c >
1$. The Factoring game is naturally asymmetric. First, it is
sequential: Alice chooses an number, and then Bob acts based on
knowledge of Alice's number. Also, the natural application of
the Factoring game is to cryptography, with Alice using a
cryptosystem and Bob trying to break it. In this context, by
the polynomial-time Church-Turing thesis, the computational
model Bob uses is at most polynomially faster than that of
Alice.

Note that analogues of our results also go through for other
dependences of $\delta$ on $\epsilon$. The choice we make is
partly intended to illustrate that our model can capture one of
the typical assumptions of complexity-theoretic cryptography.

% Thus, to make the game fair, it is natural to choose
%$\delta < \epsilon$ - the precise dependence is less important, and
%we discuss this at the end of this section.

We first show that if Factoring is easy in the worst case, then
every uniform NE of the discounted game yields payoff 2 to Bob.

\begin{theorem}
\label{Player2Win} If for all linear-time samplable
distributions $D$, Factoring can be solved in probabilistic
polynomial time with success probability $1-o(1)$ over $D$,
then for all sufficiently large $c$, the $(\epsilon,
\epsilon^c)$-discounted version of the Factoring game has a
uniform Nash equilibrium with payoff $(1,2)$, and $(1,2)$ is
the only uniform equilibrium payoff.
\end{theorem}

This result follows from the following lemma, which gives a
tighter connection between the feasibility of Factoring and the
uniform equilibrium payoffs of the discounted game.

\begin{lemma}
\label{EasyCase} If, for all linear-time samplable
distributions $D$. Factoring can be solved in probabilistic
time $o(n^c)$ with success probability $1-o(1)$ over $D$, then
there is a uniform Nash equilibrium of the $(\epsilon,
\epsilon^c)$-discounted version of the Factoring game yielding
a payoff of $(1,2)$. Moreover, if $c > 1$, then {\it every}
uniform equilibrium yields payoff $(1,2)$.
\end{lemma}

\begin{proof}
We first show the existence of the claimed uniform equilibrium
giving a payoff of $(1,2)$, and then show that this is the only
uniform equilibrium payoff achievable.

The following pair of probabilistic machines $(M, N)$ gives a
pure-strategy uniform equilibrium with payoff $(1,2)$. $M$
simply outputs the number 2. $N$ uses the trivial deterministic
algorithm for Factoring running in exponential time to find a
prime factorization for the number produced by $M$.

As $\epsilon \rightarrow 0$, the payoff for this pair of
strategies tends to $(1,2)$. We now show that $(M,N)$ is a
uniform NE for the game.

Since the payoff for Bob is bounded above by $2$, irrespective
of what it does, it's clear that the advantage it can gain from
playing a different strategy tends to zero as $\epsilon$ tends
to zero. We still need to show that Alice can't do any better
in the limit.

Let $S$ be any (mixed) strategy for Alice - $S$ is a
probability distribution over probabilistic TMs. Whenever $S$
outputs a number, player 1 gets payoff at most 1, since Bob
factors the number. When $S$ does not output a number, player 1
gets payoff 0; thus, in either case, Alice's payoff is at most
1. This shows that Alice can't do better than playing $M$.

Showing that $(1,2)$ is the only uniform equilibrium payoff
possible is more involved. For the purpose of contradiction,
let $(a,b)$ be a uniform equilibrium payoff, where either $a
\neq 1$ or $b \neq 2$. We derive a contradiction.

We first consider the case $a \neq 1$. It cannot be the case
that $a < 1$, since Alice can always get payoff at least 1 in
the limit by just outputting 1, irrespective of what Bob does.
Thus it must be the case that $1 < a \leq 2$.

Now we show that $b = 2$. Let $(S,T)$ be a uniform NE with
payoff $(a,b)$. Let $\gamma(\epsilon)$ be the probability that
$S$ outputs a number with length at most $1/\epsilon$, where
the probability is over the randomness of choosing a strategy,
as well as the randomness in playing one (since a pure strategy
is a probabilistic TM). We show that $\gamma(\epsilon)
\rightarrow 1$ as $\epsilon \rightarrow 0$. For the sake of
contradiction, suppose that the limit infimum of
$\gamma(\epsilon)$ is less than $\alpha < 1$. This means that
we can choose $\epsilon$ arbitrarily small for which $S$
outputs a number with length at least $1/\epsilon$ with
probability at least $1-\alpha$. Conditioned on outputting such
a number, the payoff of Alice is at most
$2(1-\epsilon)^{1/\epsilon}$ which tends to $2/e < 1$ as
$\epsilon \rightarrow 0$. From the previous para, we know that
Alice gets payoff at least 1 from playing $S$, hence from an
averaging argument, we can choose $\epsilon$ arbitrarily small
for which there is a probability $\beta$ bounded away from 0
that Alice outputs a number of length at most $1/\epsilon$ and
gets a payoff greater than 1. We show that in this case, Bob
can improve its payoff by a non-trivial amount by playing a
different strategy $T'$.

When defining $T'$, we use the assumption that Factoring is
easy on average for all linear-time samplable distributions
(note that this assumption was not used in the argument that
there's a uniform equilibrium with payoff $(1,2)$). Consider
the linear-time samplable distribution $D$ on inputs of length
$|1/\epsilon|$ defined as follows: Simulate $S$ independently
$k/\beta$ times for $1/\epsilon$ computation steps (where $k$
is a constant to be decided later), and output the first number
produced by $S$ of length at most $|1/\epsilon|$, padded up to
length $|1/\epsilon|$, outputting an arbitrary number of that
length if all the runs of $S$ give numbers that are too long.
Clearly $D$ is linear-time samplable. Here we use the fact that
Factoring is {\it paddable} to any given length (padding here
just involves multiplication by a power of two). There is some
algorithm $A$ that works with success probability $1-o(1)$ over
$D$, by assumption.

Consider the following strategy $T'$ for Bob: it looks at the
number output by Alice. If this number is at most $1/\epsilon$
bits long, it applies $A$ to this number. If the number is
longer, it plays strategy $T$. The process of looking at the
number and deciding what to do based on its length takes time
$O(1/\epsilon)$, but if $c > 1$, then $( 1 -
\epsilon^c)^{O(1/\epsilon)} \rightarrow 1$ when $\epsilon
\rightarrow 0$, and hence this additive term incurs a
negligible discount for Bob. Conditioned on Alice outputting a
number that's at least $1/\epsilon$ bits long, Bob's payoff is
the same in the limit when playing strategy $T'$ as when
playing strategy $T$. In the other case, Bob gets a payoff of
at least $2(1-e^{-k})$ in the limit (since he successfully
factors while using $(o(1/\delta)$ time), which for large
enough $k$ is strictly better than it did when playing strategy
$T$, given our assumption that Alice had a probability bounded
away from 0 of outputting a number at most $1/\epsilon$ bits
long and getting a payoff greater than 1 (which would imply Bob
got a payoff less than 2). This is a contradiction to $(S,T)$
being a uniform NE.

Thus, we get that $\gamma(\epsilon) \rightarrow 1$ as $\epsilon
\rightarrow 0$. But then the strategy of Bob which simply
applies the $o(n^c)$ factoring algorithm to the number output
by Alice gets a payoff of 2 in the limit. This implies that
$b=2$.

If $a \geq 1$ and $b=2$, it must be the case that $(a,b) =
(1,2)$ for the uniform NE $(S,T)$, since the expected payoff of
any pair of strategies in this game is bounded above by $3$.
\end{proof}

Next, we show an essentially converse. If Factoring is hard on
average, then there is a uniform NE for the discounted game
with payoff $(2,1)$.

\begin{theorem}
\label{Player1Win} Suppose there is a linear-time samplable
distribution $D$ for which there is no probabilistic polynomial
time algorithm correctly factoring with success probability
$\Omega(1)$ over $D$ on inputs of length $n$ for infinitely
many $n$. Then for every constant $c \geq 1$, there is a
uniform NE for the $(\epsilon, \epsilon^c)$-discounted version
of the Factoring game with payoff $(2,1)$.
\end{theorem}

The key to the proof of Theorem~\ref{Player1Win} is in the
following Lemma~\ref{HardCase} which, similar to above, makes a
stronger connection between $c$ and the running time of a
factoring algorithm. The uniform NE which we show to exist is a
simple one where Alice plays a random number of length
approximately $1/\epsilon$ and Bob halts immediately without
output. We show that any deviating strategy for Bob which gets
him an improved payoff in the limit can be transformed into a
probabilistic polynomial-time algorithm which factors well on
average.

\begin{lemma}
\label{HardCase} Suppose there is no algorithm for factoring
running in time $n^c \polylog(n)$ for large enough input length
$n$, and succeeding on a $\Omega(1)$ fraction of inputs for
infinitely many input lengths $n$. Then there is a uniform NE
for the $(\epsilon, \epsilon^c)$-discounted version of the
Factoring game with payoff $(2,1)$.
\end{lemma}

\begin{proof}
The following pair of strategies $(M,N)$ is a uniform NE. $M$
selects a number of length $n(\epsilon) = \lceil 1/\epsilon
\rceil \lceil 1/\log(\lceil 1/\epsilon \rceil) \rceil$ at
random and outputs the number. $N$ halts immediately without
output.

First we show that this gives payoff $(2,1)$. It's clear that
the payoff for $N$ is 1 since it halts without output.
Therefore the undiscounted payoff for $M$ is 2. We show that
the discounting makes a negligible difference to this, since
$M$ doesn't need to spend too much time generating a random
number of length $n(\epsilon)$. Specifically, given the number
$\lceil 1/\epsilon \rceil$ on its input tape, $M$ computes
$n(\epsilon)$ in unary and stores it on a separate tape - this
can be done in time $O(n(\epsilon))$. It then generates a
random number on the output tape, using the computed value of
$n(\epsilon)$ to ensure the number is of the right length. The
total time taken by $M$ is $O(n(\epsilon)) = O(1/(\epsilon
\log(1/\epsilon)))$, and the discounting due to this is $(1 -
\epsilon)^{O(n(\epsilon))}$, which is 1 in the limit as
$\epsilon \rightarrow 0$.

Next we show $(M,N)$ is a uniform NE. Alice has payoff bounded
above by 2 for any strategy it plays, so clearly it cannot do
better with a different strategy $S$. The bulk of the work is
showing that Bob cannot do better.

Suppose, on the contrary that there is a strategy $T$ for Bob
such that the strategy pair $(M,T)$ yields payoff at least
$1+\gamma$ for Bob for arbitrarily small $\epsilon$, where
$\gamma > 0$. We show how to extract from $T$ an algorithm that
factors efficiently on average on infinitely many input
lengths.

Choose a infinite sequence $\epsilon_1, \epsilon_2 \ldots$ such
that for each $i, 1 \leq i \leq \infty$, the strategy pair
$(M,T)$ yields payoff at least $1 + \gamma$ for Bob in the
$(\epsilon_i, \epsilon_i^c)$-discounted game, and
$n(\epsilon_i) > n(\epsilon_{i-1})$. Such a sequence exists by
the assumption that $(M,N)$ is not a uniform NE.

We show that there must exist a pure strategy $N$ for Bob such
that there is an infinite set $I$ for which $(M,N)$ yields
payoff at least $1 + \gamma/2$ for Bob for all $\epsilon_i$
such that $i \in I$. This argument takes advantage of the fact
that the Factoring game has payoffs bounded above by 2. By a
Markov argument, it must be the case for each $i \in
\mathbb{N}$ that the pure strategies in the support of $T$
which yield payoff at least $1 + \gamma/2$ must have
probability weight at least $ \gamma/2$. Now, if each pure
strategy only yields payoff at least $1 + \gamma/2$ finitely
often, then we can choose $i$ large enough so that the pure
strategies yielding payoff at least $1 + \gamma/2$ in the
$(\epsilon_i, \epsilon_i^c)$-discounted game have probability
weight less than $\gamma/2$ in the support of the mixed
strategy $T$, which is a contradiction.

Let $B = \{n(\epsilon_i), i = 1 \in I\}$.  $B$ is an infinite
set, by assumption.

We use $N$ to define a probabilistic algorithm $A$ for solving
Factoring well on average on all large enough input lengths in
$B$, contradicting the assumption of the theorem. Given an
number $x$ of length $n$, $A$ simply runs $N$ on $x$ $\log(n)$
times independently, halting each run after time $n^c \lceil
\log(n)^{c+2} \rceil$. If any of these runs outputs numbers
$y_1$ and $y_2$ such that $y_1 * y_2 = x$, $A$ outputs these
numbers, otherwise it outputs nothing. The running time of $A$
is $O(n^c \log(n)^{c+3})$. We prove that for at least an
$\Omega(1)$ fraction of strings of length $n$ for infinitely
many $n$, $A$ factors correctly with probability $1 - o(1)$.

The idea is to analyze the payoff for Bob from the strategy
$(M,N)$, and show that an expected payoff greater than 1 means
that a significant fraction of computation paths must halt
quickly and factor correctly. Given a number $x$ of length
$n(\epsilon) \in B$ and a computation path $z$ of $N$ when
given $x$, let $I_{xz} = 1$ if path $z$ terminates in a correct
factoring of $x$ and 0 otherwise, $t_{xz}$ be the time taken
along path $z$, and $p_{xz}$ be the probability of taking path
$z$. We have that, for any $x$, $\Sigma_z p_{xz} = 1$. Let
$f(x) = \Sigma_{z} (1 + I_{xz})p_{xz}(1-\delta)^{t_{xz}}$,
where $\delta = \epsilon^c$. Then the payoff of Bob is
$\Sigma_{x} f(x)/2^{n(\epsilon)}$. By assumption, this quantity
is at least $1 + \gamma/2$. By a Markov argument, this implies
that for at least a $\gamma/4$ fraction of strings $x$ of
length $n(\epsilon)$, $f(x) \geq 1 + \gamma/4$.

Fix any such $x$. We classify the computation paths $z$ for the
computation of $N$ on $x$ into three classes. The first is the
set of $z$ for which $I_{xz} = 0$. This set contributes at most
$\Sigma_{z} p_{xz}(1-\delta)^{t_{xz}} \leq \Sigma_{z} p_{xz}
\leq 1$ to $f(x)$. The next class is the set of $z$ for which
$I_{xz} = 1$ and $t_{xz} \geq 2 \log(1/\delta)/\delta$. This
set contributes at most $\Sigma_{z} 2p_{xz}(1-\delta)^{2
\log(1/\delta)/\delta} \leq \Sigma_{z} 2p_{xz}\delta \leq 2
\delta = o(1)$ to $f(x)$ (here the $o(1)$ refers to dependence
on $n(\epsilon)$ as $\epsilon \rightarrow 0$). Thus we have
that $\Sigma_{z \in Z} p_{xz} \geq \gamma/4 - o(1)$, where $Z$
is the set of $z$ for which $I_{xz}= 1$ and $t_{xz} < 2
\log(1/\delta)/\delta$.

This means that with probability at least $\gamma/4$ over
strings $x$ of size $n(\epsilon) \in B$, $N$ halts in time at
most $2 \log(1/\delta)/\delta$ and outputs factors of $x$ with
probability at least $\gamma/4 - o(1)$. This implies that for
all large enough $n \in B$, with probability at least $\gamma/4
- o(1)$ over numbers of size $n$, $N$ halts in time at most
$n^c \lceil \log(n)^{c+2} \rceil$ and factors $x$ with
probability at least $\gamma/4 - o(1)$ (we're simply upper
bounding the time as a function of $n$ rather than of
$\delta$).

Since $A$ amplifies the success probability of $N$ by running
it $\log(n)$ times independently, the success probability of
$A$ is at least $1-o(1)$ on a $\Omega(1)$ fraction of inputs,
for infinitely many input lengths.
\end{proof}

Essentially the same proof gives a more general version of
Lemma~\ref{HardCase} - if there is {\it some} linear-time
samplable distribution $D$ such that no probabilistic algorithm
running in time $n^c \polylog(n)$ achieves an $\Omega(1)$
success probability for Factoring over $D$, then there is a
uniform Nash equilibrium for the $(\epsilon,
\epsilon^c)$-discounted Factoring game achieving a limit payoff
of $(2,1)$. The only difference is that $M$ plays a random
number selected according to $D$, and we argue with respect to
this distribution rather than with respect to the uniform
distribution when defining the factoring algorithm $A$.
Theorem~\ref{Player1Win} follows immediately from this more
general version.

Unlike in the case of Lemma~\ref{EasyCase}, this is not the
only uniform Nash equilibrium when Factoring is hard. Indeed,
an examination of the proof of Lemma~\ref{EasyCase} shows that
we did not actually use the assumption when showing there was a
uniform NE with payoff $(1,2)$; the assumption was only to
prove the second part of the theorem. Thus, even when Factoring
is hard, there is a uniform NE with payoff $(1,2)$.

However, an important point to note is that the discounted
Factoring game is a  sequential game, where Alice {\it plays
first}. Thus, even though there might be a uniform NE with
payoff $(1,2)$, Alice can control which Nash equilibrium is
reached, and it is natural for it to select the equilibrium
giving it a higher payoff. The key question in the discounted
Factoring game is whether there {\it exists} a uniform NE
giving Alice a payoff greater than 1 - Lemma~\ref{EasyCase}
shows that when Factoring is easy, there isn't, and
Lemma~\ref{HardCase} shows that when Factoring is hard on
average, there is. This is somewhat related to the notion of
{\it subgame-perfect} equilibria in traditional game theory
\cite{LeytonBrown-Shoham08}. It's an interesting challenge to
define an appropriate notion of subgame-perfection for our
model which could also be used in a variation of our model
where both Alice and Bob are discounted by the total time taken
by the two of them.

If one interprets Alice getting a payoff higher than 1 as
Player 1 ``winning'' the game, this result is in close
accordance with intuition. Alice wins the game if and only if
Factoring is hard. In practice, Factoring is believed to be
hard, and therefore in practice, we expect Alice to win the
game, and not Bob as traditional game theory would predict.

The uniform equilibrium in the statement of
Lemma~\ref{EasyCase} yielding a payoff of $(1,2)$ is in fact
also a {\it strong} uniform equilibrium - this follows easily
from the proof. Can Alice hope for a strong uniform equilibrium
yielding it a payoff of 2 in the case that Factoring is hard?
The answer is no.

\begin{theorem}
\label{NonStrongUniform} Consider the $(\epsilon, \delta)$
discounted version of Factoring, where $\delta = o(\epsilon)$.
Let $(S,T)$ be any strong uniform NE of this game.Then the
payoff pair corresponding to $(S,T)$ is $(1,2)$.
\end{theorem}

\begin{proof}
The proof is very similar to the proof of the second part of
Lemma~\ref{EasyCase}, except that we can no longer use the
assumption that Factoring is in polynomial time. But we can use
an alternate strategy $N_{\epsilon}$ for Bob which plays the
role of the factoring algorithm in the proof of
Lemma~\ref{EasyCase}.

$N_{\epsilon}$ simply implements a look-up table, which stores
the numbers which $S$ may output, along with their factors.
$N_{\epsilon}$ need only store numbers of length $1/\epsilon$,
together with their factors. The key is that just by encoding
the look-up table in its state machine, $N_{\epsilon}$ can find
the factors of the number output by $S$ in time
$O(1/\epsilon)$, and since $\delta =o(\epsilon)$, this means
that the discount factor is 1 in the limit. The rest of the
argument is the same as in the proof of the second part of
Theorem \ref{EasyCase}.
\end{proof}

Of course the dependence of the strategy of Bob on $\epsilon$
is essential, since we know that there is a uniform equilibrium
yielding Alice a payoff of 2 in the limit. Moreover, the proof
illustrates why the notion of a strong uniform NE might be too
strong an equilibrium concept - Bob can push Alice's limit
payoff down to 1, but the proof involves it playing strategies
whose sizes grow exponentially in $1/\epsilon$! For small
values of $\epsilon$, this is clearly infeasible.

The issue here is that there is a tradeoff between hardware and
time. Computations can be made very efficient by exponentially
increasing hardware, but in the physical world, both hardware
and time are costly. Our model explicitly captures the idea of
time being costly through discounting, but the expense of
hardware is captured implicitly in the uniform equilibrium
concept.

There are other ways of defining equilibrium concepts which can
capture the cost of hardware in a more explicit manner. For
instance, we could define an $f(\epsilon, \delta)$-resilient
uniform NE as a uniform NE where no player gains in the limit
by playing a pure strategy whose size is bounded by
$f(\epsilon, \delta)$. Since a pure strategy is just a
probabilistic Turing machine, ``size'' has a natural
representation - it's the number of bits required to explicitly
present the state space, transition function and alphabet of
the Turing machine. A uniform NE as we define it an
$O(1)$-resilient uniform NE, while strong uniform NE are
$f(\epsilon, \delta)$-resilient uniform NE for $f$ arbitrarily
large.

Now let us consider $f(\epsilon, \delta)$-resilient NE where
$\delta$ is polynomially bounded in $\epsilon$, and $f$ is
polynomially bounded in $1/\epsilon$. By using essentially the
proof of Theorem \ref{HardCase}, as well as the fact that a
probabilistic Turing machine of size K and operating in time T
can be simulated by a probabilistic Boolean circuit of size
$O(K+T)^2$, we get that there there is an $f(\epsilon,
\delta)$-resilient uniform NE giving Alice a payoff of 2 in the
limit, unless Factoring can be solved correctly by
polynomial-size circuits on an $\Omega(1)$ fraction of inputs,
for large enough input lengths.

Thus, not only does is the difference between feasibility and
infeasibility of factoring captured by a difference in the
structure of equilibria for the Factoring game, but by a
natural modification of the notion of uniformity, we can
capture the difference between uniformity and non-uniformity!
This raises the possibility that there might be interesting
{\it concrete complexity} notions that might be captured by
game theory as well - we need not restrict attention to what
happens in the limit as $\epsilon \rightarrow 0$. Perhaps there
are novel notions of complexity that can be extracted from the
game-theoretic viewpoint, which give a better understanding of
the gap between finite complexity and asymptotic complexity?

We conclude this section by discussing our choice of parameters
for the Factoring game, and showing that the results are robust
to the choices we make. First, we examine the payoffs. Any
choice of payoffs which are all positive and for which Bob gets
strictly more (resp. Alice gets strictly less) if Bob succeeds
in factoring will yield essentially equivalent results.

Second, we discuss the discount factors. Our choice of
dependence of $\delta$ on $\epsilon$ was made to illustrate
nicely the correspondence between infeasibility and the
existence of equilibria yielding Alice a high payoff. But the
polynomiality of the dependence is not critical to our proofs -
in general, if $1/\delta = f(1/\epsilon)$ for some function
$f$, then our results hold when feasibility means solvability
in time $o(f(n))$ and infeasibility means unsolvability on
average in time slightly more than $f(n)$.

In the special case that $\delta = \epsilon$, we get that Alice
has a winning strategy under the natural assumption that
Factoring is not in quasi-linear time on average.

\section{Properties of Discounted Time Games}
\label{properties}

The most fundamental results in a theory of games of a given
form concern existence of equilibria. Here we prove a couple of
results of this form. The first result shows that the concept
of uniform equilibrium for the discounted version of a finite
game corresponds nicely to the concept of Nash equilibrium for
the original game. The second result complements this by
showing that discounted games might have equilibria that the
original game does not possess.

We show that any Nash equilibrium in a finite game $G$
translates to a strong uniform Nash equilibrium yielding the
same uniform payoff in the discounted version of $G$.

\begin{theorem}
\label{finite} Let $G$ be a finite two-player game. Given any
Nash equilibrium $(S,T)$ of $G$, there is a strong uniform Nash
equilibrium $(S',T')$ of the discounted version of $G$ which
yields the same payoff in the limit as $\epsilon, \delta
\rightarrow 0$.
\end{theorem}

\begin{proof}
We assume that $G$ is a finite two-player game in normal form.
If $G$ is sequential and given in extensive form, we just
consider the image normal-form game, which is known to inherit
its equilibria from the sequential game.

Let $(S,T)$ be a (possibly mixed-strategy) NE of $G$. We define
a strategy pair $(S',T')$ for the discounted version of $G$,
and argue that this is a strong uniform Nash equilibrium for
the discounted version, with the same payoffs for both players
in the limit. Given any pure strategy $s_1$ of a player in $G$,
choose in an arbitrary way a Turing machine $M_{s_1}$ which
ignores its input and halts after outputting a representation
of $s_1$. If $S$ gives probability $p_1$ to strategy $s_1$,
then we give machine $M_{s_1}$ probability $p_1$ in $S'$. $T'$
is defined in an analogous way given $T$.

The key point is that irrespective of the way the
representative machines for strategies are chosen, they are
guaranteed to halt in finite time. As $\delta$ and $\epsilon$
approach zero, the discount factors approach one, and hence the
payoff in the discounted game from playing $(S',T')$ approaches
the payoff from playing $(S,T)$ in $G$.

It still remains to be shown that $(S',T')$ is an $\eta$-NE for
the discounted game, where $\eta \rightarrow 0$ when $\epsilon,
\delta \rightarrow 0$. This would imply that $(S',T')$ is a
strong uniform NE for the discounted game. We show that player
1 cannot gain a significant advantage from playing a different
mixed strategy $S'_1$ - the analogous result holds for Player 2
as well.

Any mixed strategy $S'_1$ in the discounted game can be
transformed into a mixed strategy $S_1$ in $G$ - each pure
strategy is given the same probability of being played in $G$
as it has of being output by a probabilistic TM in the
discounted game (the probability weight of non-halting
computation paths is assigned to an arbitrary strategy in
$S_1$). Because of the discounting, the payoff that Player 1
can get by playing $S'_1$ in the discounted game is at most the
payoff that he can get by playing $S_1$ in $G$. But the payoff
by playing $S$ in $G$ is at least the payoff by playing $S_1$
in $G$, and the payoff by playing $S'$ in the discounted game
approaches the payoff by playing $S$ in $G$ as $\epsilon,
\delta \rightarrow 0$. This shows that the advantage of playing
$S'_1$ in the discounted game must tend to zero as $\epsilon,
\delta$ tend to zero, for an arbitrary $S'_1$, implying that
$(S',T')$ is a strong uniform NE for the discounted game.
\end{proof}

Consider the Largest Integer Game where both players
simultaneously play integers. The player playing the largest
integer receives a payoff of 100 with each receiving 50 if they
play the same integer. This game has no Nash equilibrium or
even an almost Nash equilibrium (Nash's theorem doesn't apply
because the action space is not compact).

Next we show that almost-NEs exist, not only for the Largest
Integer game for but any countable game with bounded payoffs.
The basic idea of the proof is to approximate the discounted
countable game by a finite game, and then reduce the existence
of uniform equilibria in the discounted countable game to the
existence of NEs in the corresponding finite game.

\begin{theorem}
\label{EqExist} Let $G$ be a two-player game with bounded
payoffs where both players have a countable number of actions.
Then for each $\epsilon, \delta
> 0$, the $(\epsilon, \delta)$-discounted time version of $G$ has
an $(\epsilon + \delta)$-NE.
\end{theorem}

\begin{proof}
Let $G$ be as stated in the theorem, and let $K \geq 1$ be an
upper bound on payoffs for $G$. Consider the $(\epsilon,
\delta)$-discounted time version of $G$. We show how to
approximate the discounted game by a finite game $G_{\epsilon,
\delta}$ and then use the existence of Nash equilibria in the
finite game to show the existence of approximate Nash
equilibria in the discounted game.

The finite game $G_{\epsilon, \delta}$ is the subgame of the
discounted game where the first player plays probabilistic
Turing machines of description size at most
$2^{2K^2/\epsilon^2}$, and the second player plays
probabilistic Turing machines of size at most
$2^{2K^2/\delta^2}$. By Nash's theorem, this game has a
mixed-strategy Nash equilibrium $(S_1,T_1)$. We show that
$(S_1,T_1)$ is an $(\epsilon+\delta)$-NE for the discounted
game.

We show that for {\it any} mixed strategy pair $(S,T)$ in the
discounted game, there is a mixed strategy pair $(S',T')$ in
$G_{\epsilon,\delta}$ such that $u_2(S, T') \geq P_2(S,T) -
\delta$, and $u_1(S',T) \geq P_1(S,T) - \epsilon$. This implies
that any NE for $G_{\epsilon, \delta}$ is an
$(\epsilon+\delta)$-NE for the discounted game.

Let $(S,T)$ be a mixed strategy pair in the discounted game. We
show how to construct a strategy $T'$ in $G_{\epsilon, \delta}$
for Player 2 such that $u_2(S, T') \geq u_2(S,T) - \delta$. The
corresponding result for Player 1 follows by a symmetric
argument.

The argument is a ``probability-shifting'' argument - we will
show how to transfer probability from probabilistic machines in
the support of $T$ with size more than $2^{2K^2/\delta^2}$ to
probabilistic machines with description size smaller than that
number without damaging the payoff of Player 2 too much.
Specifically, the payoff of Player 2 will not decrease by more
than $\delta$ conditional on that strategy being played, and
hence there will not be more than a $\delta$ decrease in total.

Let $N$ be a probabilistic machine of size more than
$2^{2K^2/\delta^2}$ which has non-zero weight in $T$. We define
a corresponding machine $N'$ of size at most
$2^{2K^2/\delta^2}$, and transfer all the probability weight of
$N$ to $N'$ in $T'$. Essentially, $N'$ will be
indistinguishable from $N$ relative to the discounting.

The key observation is that we don't need to take into account
computation paths in $N$ of length greater than $K^2/\delta^2$,
because the strategies output on such computation paths are so
radically discounted that we may as well assume they yield zero
payoff, without incurring too much damage to the overall
payoff. $N'$ behaves like $N$ ``truncated'' to $K^2/\delta^2$
steps, outputting a strategy for $G$ if $N$ does within that
time, and looping otherwise.

We cannot simply simulate $N$ using a universal machine and a
clock, since the simulation takes too much of a time overhead
and does not preserve the payoff to within a small additive
overhead. Instead we simulate $N$ in {\it hardware} - this is
much more time efficient. Specifically, we're interested in the
behavior of $N$ only for the first $K^2/\delta^2$ time steps.
We can define a Turing machine $N'$ with description size at
most $2^{2K^2/\delta^2}$ which encodes the relevant behavior of
$N'$ entirely in its finite state control. This simulation
incurs no time overhead at all.

Now, we calculate the maximum damage to Player 2's payoff from
playing $N'$ instead of $N$. There is no damage to the payoff
from computation paths of $N$ which terminate within
$K^2/\delta^2$ steps. Thus the loss in payoff is bounded above
by $(1-\delta)^{K^2/\delta^2}K$, which is at most $\delta$ if
$K \geq 1$. This finishes the argument.
\end{proof}

% Took out to save space -- Lance
%Theorem \ref{EqExist} is very general - there is no restriction
%on the computability of payoffs, for instance. Because of the
%generality of the assumption, we cannot say much about the
%uniformity or otherwise of the NEs for the discounted game.

In case the payoffs of the game $G$ are computable, we get a
stronger version of Theorem \ref{EqExist} in that uniform
equilibria are guaranteed to exist.

\begin{theorem}
\label{UniformExist} Let $G$ be a two-player game where each
player has a countable number of actions, and suppose the
payoffs are bounded and computable. Then the discounted time
version of $G$ has a uniform equilibrium.
\end{theorem}

\begin{proofsketch}
The proof is similar to the proof of Theorem \ref{EqExist}, but
we take advantage of the fact that payoffs are computable. As
in the proof of Theorem \ref{EqExist}, we can define a finite
truncated version of the game such that the almost-Nash
equilibria of the truncated game are also almost-Nash
equilibria of the discounted game. In order to ensure
uniformity, however, we have to produce a {\it fixed} pair of
strategies such that as $\epsilon, \delta \rightarrow 0$,
neither can gain a non-zero amount in the limit by using a
different strategy.

The basic idea is to define a strategy pair $(M,N)$ such that
$M$ and $N$ deterministically compute an almost-Nash
equilibrium of the truncated game, with $M$ proceeding to play
the strategy of player 1 in the computed almost-Nash
equilibrium, and $N$ proceeding to play the strategy of player
2. There are two obstacles to this approach. The first is the
computational obstacle, but this can be circumvented since the
entries of the payoff matrix for the truncated game can be
estimated to any desired accuracy using sampling and the
computability of the payoffs of the original game, and then the
Lemke-Howson algorithm \cite{Lemke-Howson64} can be used to
find almost-equilibria of the truncated game.

The second obstacle is that computing an almost-Nash
equilibrium of the truncated game incurs a substantial time
overhead, which already drives the payoffs of the two players
down before they play the strategies corresponding to the
almost-Nash equilibrium, not to mention the simulation overhead
from using a single machine ($M$ or $N$) to find an almost-Nash
equilibrium for all $\epsilon, \delta > 0$. This obstacle is
overcome using the idea of ``miniaturization'' - given discount
rates $\epsilon$ and $\delta$ respectively, the players pretend
that their discount rates are $\epsilon'$ and $\delta'$
instead, where $1/\epsilon'$ and $1/\delta'$ grow very slowly
as a function of $1/\epsilon$ and $1/\delta$. $\epsilon'$ and
$\delta'$ are chosen so that the players can compute an
almost-Nash equilibrium of the $(\epsilon',
\delta')$-discounted game quickly enough that their payoffs in
the $(\epsilon, \delta)$-discounted game are hardly affected by
this computation, and that playing the strategies for the
truncated game takes relatively little time as well. The point
is that this is still an $(\epsilon'+\delta')$-NE for the
discounted game, and that $\epsilon', \delta' \rightarrow 0$ as
$\epsilon, \delta \rightarrow 0$. Hence it is a uniform Nash
equilibrium.
\end{proofsketch}

The bounded-payoff assumption in Theorems \ref{EqExist} and
\ref{UniformExist} is essential for the conclusion to hold.
Indeed, consider the two-player game where Player 1 derives a
payoff of $2^{i}$ from playing integer $i$ and Player 2 a
payoff of $2^{j}$ from playing integer $j$. It is not hard to
see that this game does not even have almost-NEs in the
discounted game.

%Theorem \ref{EqExist} is especially interesting for a game like the
%Largest Integer game, which has bounded payoffs but no NE or
%almost-NE. The Largest Integer game is a two-player game where both
%players play numbers. If Player 1's number is greater than Player
%2's, Player 1 gets payoff 100 and Player 2 gets payoff 0, and
%conversely. If their numbers are equal, then Player 1 and Player 2
%both get payoff 50.

Theorem \ref{EqExist} shows that the discounted version of the
Largest Integer Game does have almost-NEs. For the Largest
Integer game, in fact, there is a strong uniform equilibrium
which yields a payoff of 0 for both players, and every uniform
equilibrium gives payoff 0 to both players in the limit. This
is intuitive: the Largest Integer game is a game of
oneupmanship, where each player tries to outdo the other by
producing a larger number. In the process, they exhaust their
computational resources (or alternatively, end up spending an
inordinate amount of time) and end up with nothing.

In general, uniform equilibrium is a strong notion of
equilibrium, since there should be no gain in deviating
irrespective of {\it how} $\epsilon, \delta \rightarrow 0$.
Suppose we know more about the relationship of $\epsilon$ and
$\delta$, say that $\delta < \epsilon^2$, i.e., Player 2 always
has more computational power. In this case there are equilibria
in which Player 2 wins, say by outputting $2(1-\epsilon)^{3/2}$
while Player 1 outputs $(1-\epsilon)^{3/2}$. This is again in
accordance with intuition - if the players are asymmetric, the
more patient/computational stronger player should win this game
(the discount rate can be seen, depending on the situation, as
either an index of patience or of computational power).

\section*{Acknowledgements}
We thank Ehud Kalai for suggesting the factoring game to us and
to Ehud and Amin Saberi for useful discussions. We also thank
the anonymous reviewers for many useful suggestions.

\bibliographystyle{unsrt}
\bibliography{papers_Lance}

\end{document}